\documentclass[fleqn,usenatbib]{mnras}

\usepackage{newtxtext,newtxmath}

\usepackage[T1]{fontenc}

\DeclareRobustCommand{\VAN}[3]{#2}
\let\VANthebibliography\thebibliography
\def\thebibliography{\DeclareRobustCommand{\VAN}[3]{##3}\VANthebibliography}

\usepackage{graphicx}	
\usepackage{amsmath}	
\usepackage{booktabs}

\newcommand\ee{\end{equation}}
\newcommand\be{\begin{equation}}
\newcommand\eea{\end{eqnarray}}
\newcommand\bea{\begin{eqnarray}}

\renewcommand\[{\left[}
\renewcommand\]{\right]}

\title[Quantifying the global "CMB tension"]{Quantifying the global "CMB tension" between the Atacama Cosmology Telescope and the Planck satellite in extended models of cosmology}

\author[E. Di Valentino et al.]{
Eleonora Di Valentino,$^{1}$\thanks{E-mail: e.divalentino@sheffield.ac.uk}
William Giar\`e,$^{2,3}$
Alessandro Melchiorri$^{4}$
and Joseph Silk$^{5,6,7}$
\\
$^{1}$School of Mathematics and Statistics, University of Sheffield, Hounsfield Road, Sheffield S3 7RH, United Kingdom\\
$^{2}$Galileo Galileo Institute for theoretical physics, Centro Nazionale INFN di Studi Avanzati, Largo Enrico Fermi 2,  I-50125, Firenze, Italy\\
$^{3}$INFN Sezione di Roma, P.le A. Moro 2, I-00185, Roma, Italy\\
$^{4}$Physics Department and INFN, Universit\`a di Roma ``La Sapienza'', Ple Aldo Moro 2, 00185, Rome, Italy\\
$^{5}$Institut d'Astrophysique de Paris (UMR7095: CNRS \& UPMC- Sorbonne Universities), F-75014, Paris, France\\
$^{6}$Department of Physics and Astronomy, The Johns Hopkins University Homewood Campus, Baltimore, MD 21218, USA\\
$^{7}$BIPAC, Department of Physics, University of Oxford, Keble Road, Oxford OX1 3RH, UK
}

\date{Accepted XXX. Received YYY; in original form ZZZ}

\pubyear{2022}

\begin{document}
\label{firstpage}
\pagerange{\pageref{firstpage}--\pageref{lastpage}}
\maketitle

\begin{abstract}
We study the global agreement between the most recent observations of the Cosmic Microwave Background temperature and polarization anisotropies angular power spectra released by the Atacama Cosmology Telescope and the Planck satellite in various cosmological models that differ by the inclusion of different combinations of additional parameters. By using the Suspiciousness statistic, we show that the global "CMB tension" between the two experiments, quantified at the Gaussian equivalent level of $\sim 2.5\,\sigma$ within the baseline $\Lambda$CDM, is reduced at the level of $1.8\sigma$ when the effective number of relativistic particles ($N_{\rm eff}$) is significantly less than the standard value, while it ranges between $2.3\,\sigma$ and $3.5\,\sigma$ in all the other extended models.
\end{abstract}




\section{Introduction}
\label{sec.introduction}

Accurate measurements of the Cosmic Microwave Background (CMB) are critical to cosmology since any proposed model of the Universe must be able to explain any feature present in this relic radiation.

Historically, the first unequivocal observation of the near perfect black-body spectrum of CMB photons and their tiny temperature fluctuations was obtained in 1989 by the COBE satellite~\citep{Fixsen:1993rd, Bennett:1996ce}, opening the so-called epoch of precision cosmology. Since then, significant efforts have been devoted to improve the experimental accuracy and substantially better measurements of the CMB angular power spectra of temperature and polarization anisotropies have been released first by the WMAP satellite~\citep{Hinshaw:2012aka,WMAP:2012fli} and, more recently, by the Planck satellite~\citep{Planck:2019nip,Planck:2018nkj}, the Atacama Cosmology Telescope (ACT-DR4)~\citep{ACT:2020frw,ACT:2020gnv} and the South Pole Telescope (SPT)~\citep{SPT-3G:2014dbx,SPT-3G:2021eoc}.

All these independent CMB experiments are broadly in agreement with a vanilla $\Lambda$CDM model of structure formation that, along the years, has been established as the standard concordance model of cosmology. It describes a spatially flat Universe dominated at late times by a cosmological constant $\Lambda$ and in which the majority of matter interacts only gravitationally. We call this type of matter cold dark matter (CDM) and parameterize it as a perfect fluid of collisionless particles. In addition, to set the appropriate initial condition, we introduce an early phase of cosmological inflation~\citep{Guth:1980zm} that is supposed to drive the Universe towards spatial flatness and large-scale homogeneity; providing, at the same time, a robust framework for explaining the origin of primordial density fluctuations. Last but not least, we assume that General Relativity is the correct theory of gravitation and that the other fundamental interactions obey the Standard Model of particle physics.

However, the three major unknown ingredients of the standard cosmological model (\textit{i.e.}, Inflation, Dark Matter and Dark Energy), although absolutely necessary to explain observations, still lack solid theoretical interpretations and direct experimental measurements. In this sense, $\Lambda$CDM resembles a phenomenological data-driven approximation to a more accurate scenario that has yet to be fully explored (or even understood) on a more fundamental level. Therefore, it is entirely plausible that the same model may prove inadequate to fit more precise observations from widely different cosmic epochs and scales. Interestingly, in recent years, as error-bars on cosmological parameters begun to narrow, intriguing tensions and anomalies have emerged at different statistical levels (see, \textit{e.g.}~\citep{DiValentino:2020vvd,DiValentino:2020zio,DiValentino:2020srs,Abdalla:2022yfr} and references therein).  

The most significant problem is the so-called "Hubble tension": a $5.3\sigma$ inconsistency~\citep{Riess:2022mme} between the value of the Hubble constant measured by the SH0ES collaboration using luminosity distances of Type Ia supernovae calibrated by Cepheids (see also~\citep{Riess:2021jrx} that gives $H_0=73\pm1$ km~s$^{-1}$~Mpc$^{-1}$) and the result obtained by Planck satellite~\citep{Planck:2018vyg} ($H_0=67.4\pm0.5$ km~s$^{-1}$~Mpc$^{-1}$). And while all of the early time $H_0$ estimates assuming a $\Lambda$CDM model are in agreement with Planck, all of the late-time measurements are instead in agreement with SH0ES (see~\cite{DiValentino:2021izs,DiValentino:2020zio,Perivolaropoulos:2021jda,Abdalla:2022yfr} and references therein), and the tension persists even when removing some of the measurements (see~\cite{Verde:2019ivm,Riess:2019qba,DiValentino:2020vnx,DiValentino:2022fjm}). Other minor yet relevant tensions concern the value of the clustering parameters $S_8$ and $\sigma_8$, now above $3\sigma$, inferred by CMB and weak lensing experiments~\citep{DiValentino:2020vvd,Heymans:2020gsg,KiDS:2020ghu,DES:2021vln,DES:2022ygi}, the Planck anomalous preference for a higher lensing amplitude at about $2.8$ standard deviations~\citep{Planck:2018vyg,DiValentino:2019dzu,DiValentino:2020hov} and the indication for a closed Universe at level of $3.4$ standard deviations~\citep{Planck:2018vyg,DiValentino:2019qzk,Handley:2019tkm}, often  in 
disagreement with other complementary astrophysical observables, such as Baryon Acoustic Oscillation (BAO) measurements~\citep{Beutler:2011hx,Ross:2014qpa,Alam:2016hwk} when combined with Planck.

Excitingly, these discrepancies among CMB and CMB-independent surveys may hint at new physics beyond $\Lambda$CDM, but may reflect also the presence of important observational systematics in  either or both choices of datasets. In this regard, comparing results of multiple CMB measurements is certainly a good method to  question the nature of such anomalies and discriminate between the two possibilities~\citep{DiValentino:2022oon}. For this reason, appropriate statistical methods have been developed to accurately quantify the consistency between independent CMB experiments~\citep{Charnock:2017vcd,Lin:2019zdn,Handley:2019wlz} and in~\cite{Handley:2020hdp} it  was shown that, within the $\Lambda$CDM model of cosmology,  ACT-DR4 is in mild-to-moderate tension with Planck and SPT, at a gaussian equivalent level of  2.6$\sigma$ and 2.4$\sigma$, respectively. This controversial tension,  in between $2-3\sigma$, is worthy of  being further investigated if we want to use CMB data to do "precision cosmology" and derive constraints on the fundamental physics.

In this Letter, focusing exclusively on the two most constraining CMB experiments,  we extend this analysis and quantify the global "CMB tension" between ACT-DR4 and Planck in various extended models of cosmology that differ from the baseline case by the inclusion of different combinations of additional parameters. We show that including extra degrees of freedom in the fit hardly accommodates the global tension between these two datasets, but remarkable exceptions are observed in models with less dark radiation at recombination as quantified by $N_{\rm eff}$, where the tension is reduced up to 1.8$\sigma$.

\section{Statistical Analysis}
\label{sec.methods}

\begin{table*}
\centering
\renewcommand{\arraystretch}{1.5}
\begin{tabular}{l @{\hspace{0.5 cm}} c@{\hspace{1 cm}} c @{\hspace{1 cm}} c @{\hspace{1 cm}} c @{\hspace{1 cm}} c}
\toprule
\textbf{Cosmological model} & \boldmath{$d$} & \boldmath{$\chi^2$} & \boldmath{$p$} & \boldmath{$\log S$}& \textbf{Tension}\\
\hline\hline
$\Lambda$CDM & $ 6 $ & $ 16.3 $ & $ 0.012 $ & $ -5.17 $ & $ 2.51 \,\sigma$ \\
$\Lambda\text{CDM}+A_{\text{lens}}$ & $ 7 $ & $ 18.5 $ & $ 0.00977 $ & $ -5.77 $ & $ 2.58 \,\sigma$ \\
$\Lambda\text{CDM}+N_{\text{eff}}$ & $ 7 $ & $ 13 $ & $ 0.0719 $ & $ -3 $ & $ 1.80 \,\sigma$ \\
$\Lambda\text{CDM}+\Omega_{k}$ & $ 7 $ & $ 16.5 $ & $ 0.0209 $ & $ -4.75 $ & $ 2.31 \,\sigma$ \\
$w\text{CDM}$ & $ 7 $ & $ 16.8 $ & $ 0.0187 $ & $ -4.9 $ & $ 2.35 \,\sigma$ \\
$\Lambda\text{CDM}+\sum m_{\nu}$ & $ 7 $ & $ 20.7 $ & $ 0.00421 $ & $ -6.86 $ & $ 2.86 \,\sigma$ \\
$\Lambda\text{CDM}+\alpha_s$ & $ 7 $ & $ 20.6 $ & $ 0.00448 $ & $ -6.78 $ & $ 2.84 \,\sigma$ \\
\hline
$w\text{CDM}+\Omega_{k}$ & $ 8 $ & $ 17.6 $ & $ 0.0249 $ & $ -4.78 $ & $ 2.24 \,\sigma$ \\
$\Lambda\text{CDM}+\Omega_{k}+\sum m_{\nu}$ & $ 8 $ & $ 21.2 $ & $ 0.00651 $ & $ -6.62 $ & $ 2.72 \,\sigma$ \\
$w\text{CDM}+\Omega_{k}+\sum m_{\nu}$ & $ 9 $ & $ 19.8 $ & $ 0.0195 $ & $ -5.38 $ & $ 2.34 \,\sigma$ \\
$w\text{CDM}+\Omega_{k}+\sum m_{\nu}+N_{\text{eff}}$ & $ 10 $ & $ 18.8 $ & $ 0.0434 $ & $ -4.38 $ & $ 2.02 \,\sigma$ \\
$w\text{CDM}+\Omega_{k}+\sum m_{\nu}+\alpha_s$ & $ 10 $ & $ 22 $ & $ 0.015 $ & $ -6.01 $ & $ 2.43 \,\sigma$ \\
$w\text{CDM}+\Omega_{k}+N_{\text{eff}}+\alpha_s$ & $ 10 $ & $ 20.9 $ & $ 0.0218 $ & $ -5.45 $ & $ 2.29 \,\sigma$ \\
$w\text{CDM}+\sum m_{\nu}+N_{\text{eff}}+\alpha_s$ & $ 10 $ & $ 31.1 $ & $ 0.000575 $ & $ -10.5 $ & $ 3.44 \,\sigma$ \\
$w\text{CDM}+\Omega_{k}+\sum m_{\nu}+N_{\text{eff}}+\alpha_s$ & $ 11 $ & $ 24.7 $ & $ 0.0102 $ & $ -6.83 $ & $ 2.57 \,\sigma$ \\

\bottomrule
    \end{tabular}
    \caption{Global tension between Planck and ACT-DR4 in different (extended) models of cosmology. For each model, we report the number of free parameters $d$ by~\autoref{eq:dim}, the $\chi^2$ calculated by~\autoref{eq:chi2}, the corresponding tension probability $p$ estimated by~\autoref{eq:p}, the Suspiciousness from~\autoref{eq:logS} and finally the Gaussian-equivalent tension by~\autoref{eq:sigma}. We report the minimal one-parameter extensions of the baseline $\Lambda$CDM model above the line and the higher dimensional cosmological model below the line.}
\label{tab.Results.SusStat}
\end{table*}

Aiming to quantify the global consistency between Planck and ACT-DR4 in extended models of cosmology, we start by considering the standard $\Lambda$CDM scenario described by the usual set of 6 parameters 
\be \Theta_{\Lambda\rm{CDM}} \doteq\{ \Omega_{\rm b}h^2  \, , \, \Omega_{\rm c}h^2 \, ,\theta_{\rm{MC}}\,,\, \tau \,,\,\log(10^{10}A_{\rm S}) \,,\,n_s \}\ee
and proceed by relaxing some of the underlying assumption for this baseline case, introducing  the possibilities to have a different lensing amplitude than in General Relativity ($A_{\rm lens}\ne 1$),  a number of relativistic species in the early Universe different from predictions of the Standard Model of particle physics ($N_{\rm eff}\ne 3.04$), a non-flat spacetime geometry ($\Omega_k\ne 0$), a generic Dark Energy equation of state ($w\ne -1$), massive neutrinos ($\sum m_{\nu}>0$ ) and a non-vanishing running of the scalar spectral index ($\alpha_s\equiv dn_{\rm s}/d\log k\ne 0$).  In this way, we analyze many extended cosmological models that differ for the inclusion of different combinations of $N$ additional parameters $\theta_i$
 \be \theta_i \in \{ A_{\rm lens}\,,\,N_{\rm{eff}} \,,\,\Omega_k\,,\,w\,,\,\sum m_{\nu}\,,\,\alpha_s \} \ee 
for a total number of free degrees of freedoms 
 \be d=\dim\left(\Theta_{\Lambda\rm{CDM}} \, \bigcup \,\left\{\theta_i\right\}_{i=1,\dots, N}\right)=6+N \label{eq:dim} \ee
that, in our analysis, ranges between $d=7$ and $d=11$ (\textit{i.e.}, from minimal extensions up to N=5 more parameters than $\Lambda$CDM). 

For each cosmological model, using  the publicly available package \texttt{CosmoMC}~\citep{Lewis:2002ah,Lewis:2013hha} and computing the cosmological model exploiting the latest version of the Boltzmann code \texttt{CAMB}~\citep{Lewis:1999bs,Howlett:2012mh}, we perform  a full Monte Carlo Markov Chain (MCMC) analysis of the observations of the Cosmic Microwave Background provided by Planck and ACT.  In particular, we exploit the Planck 2018 temperature and polarization (TT TE EE) likelihood~\citep{Planck:2019nip,Planck:2018vyg,Planck:2018nkj}, which also includes low multipole data ($\ell < 30$) and the Atacama Cosmology Telescope DR4 likelihood~\citep{ACT:2020frw} with a Gaussian prior on $\tau = 0.065 \pm 0.015$, as done in~\cite{ACT:2020gnv}. 

We explore the posteriors of our parameter space using the MCMC sampler developed for \texttt{CosmoMC} and tailored for parameter spaces with a speed hierarchy which also implements the "fast dragging" procedure described in~\cite{Neal:2005}. The convergence of the chains obtained with this procedure is tested using the Gelman-Rubin criterion~\citep{Gelman:1992zz} and we choose as a threshold for chain convergence $R-1 \lesssim 0.01 $. 

In order to quantify the global consistency in these extended parameter spaces and compare our results with the existing ones for the baseline case~\citep{Handley:2020hdp}, we retrace the same methodology outlined in~\citep{Handley:2019wlz,Handley:2020hdp}. In particular, we make use of the so-called Suspiciousness statistics introduced in~\citep{Handley:2019wlz} to address undesired dependencies on the prior volume. The basic idea is to divide the Bayes Ratio into two components: a prior-dependent part (the Information $I$) and a prior-independent term, the Suspiciousness $S$.  It is important to keep in mind that while ACT and Planck both independently measure the temperature and polarization angular power spectra of the Cosmic Microwave Background, the two datasets are not completely independent due to an overlap in measured multipole range (\textit{i.e.}, the two experiments are partially measuring the same sky). Therefore, to properly analyze any potential tension between these correlated datasets, the techniques outlined in \citep{Lemos:2019txn} should be used. However, currently no joint likelihood for these datasets exists. Assuming uncorrelated-datasets and gaussian-like posterior distributions for parameters with means $\mu$ and  covariance matrix $\Sigma$, the Suspiciousness can be estimated as~\cite{Handley:2019wlz,Handley:2020hdp}
\be
\log S=\frac{d}{2}-\frac{\chi^{2}}{2}
\label{eq:logS}
\ee
where the $\chi^2$ is given by
\be
\chi^{2}=\left(\mu_{A}-\mu_{B}\right)\left(\Sigma_{A}+\Sigma_{B}\right)^{-1}\left(\mu_{A}-\mu_{B}\right)
\label{eq:chi2}
\ee
with $\[A\,,\,B\]\equiv\[\rm{Planck}\,,\,\rm{ACT}\]$. Notice that the $\chi^2$ can be converted easily into a tension probability by the survival function of the $\chi^2$ distribution
\be
p=\int_{\chi^{2}}^{\infty} \frac{x^{d / 2-1} e^{-x / 2}}{2^{d / 2} \Gamma(d / 2)} d x
\label{eq:p}
\ee
and, ultimately, into a Gaussian equivalent tension via the inverse error function:
\be
\sigma(p)=\sqrt{2} \operatorname{erfc}^{-1}(1-p).
\label{eq:sigma}
\ee

Despite this procedure having a number of caveats and limitations, offering only an approximated method of assessing consistency among uncorrelated data-sets, in this work we judge it good enough to identify major inconsistencies in the two datasets. At the same time, this methodology has the benefit of providing a synthetic picture of how such inconsistencies change in extended parameter-spaces, without introducing any bias due to the prior volume effects\footnote{We note that the procedure assumes the parameter posterior distribution functions to be Gaussian distributed in such a way that what we called $\chi^2$ in~\autoref{eq:chi2}, is actually $\chi^2$--distributed. While this is of course not exactly true for some additional parameters considered in our analysis, the vast majorities of the cosmological parameters show Gaussian-like posterior distributions within a very good level of accuracy.}.

\section{Results and Discussion}
\label{sec.results}

\begin{figure*}
\centering
\includegraphics[width=0.8\textwidth]{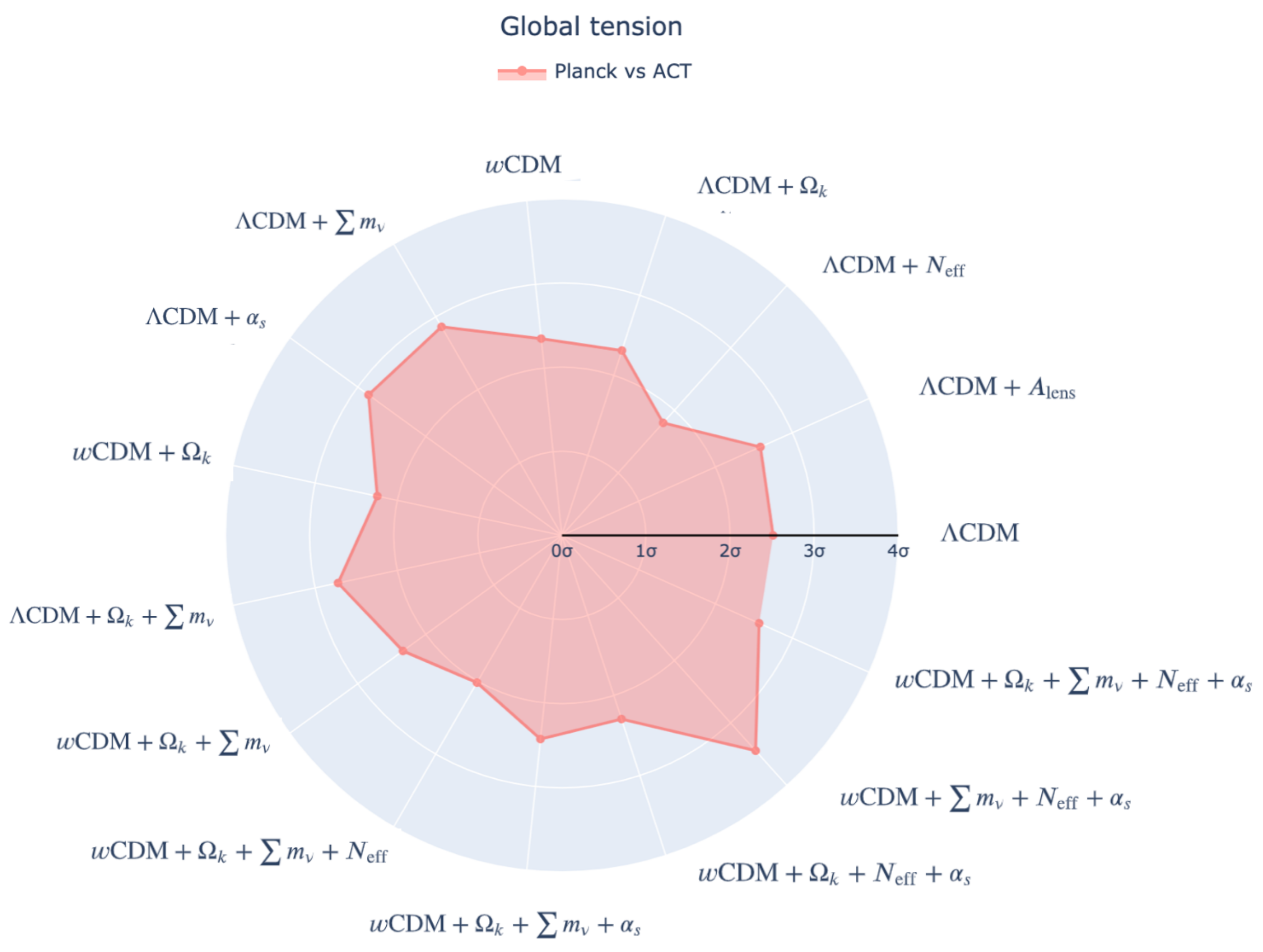}
\caption{Gaussian equivalent tension between Planck and ACT-DR4 in extended models of cosmology.}
\label{fig:sigma}
\end{figure*}

\begin{table*}
\centering
\renewcommand{\arraystretch}{1.5}
\resizebox{0.95\textwidth}{!}{\begin{tabular}{l @{\hspace{0.4 cm}} lccccccccc}
\toprule
\textbf{Cosmological model} & \textbf{Dataset} & \boldmath{$A_{\text{lens}}$} & \boldmath{$N_{\text{eff}}$} & \boldmath{$\Omega_{k}$} & \boldmath{$w$}& \boldmath{$\sum m_{\nu}$\textbf{[eV]}} & \boldmath{$\alpha_s$}\\
\hline\hline

$\Lambda\text{CDM}+A_{\text{lens}}$& Planck        &$1.180\pm 0.065$   & -- & -- &-- &-- &--\\
&ACT-DR4 &$1.01^{+0.10}_{-0.12}$   & -- & -- &-- &-- &--\\[2ex]

$\Lambda\text{CDM}+N_{\text{eff}}$ & Planck&         --   & $2.92\pm 0.19 $ & -- &-- &-- &--\\
& ACT-DR4        &--   & $2.35^{+0.40}_{-0.47}$ & -- &-- &-- &-- \\[2ex]

$\Lambda\text{CDM}+\Omega_{k}$ &  Planck&        --   & -- & $-0.044^{+0.018}_{-0.015}$ &-- &-- &--\\
 &ACT-DR4    &--   & -- & $-0.005^{+0.023}_{-0.013}$ &-- &-- &--\\[2ex]
 
$w\text{CDM}$ &  Planck&      --   & -- & -- &$-1.58^{+0.16}_{-0.35}$ &--&--\\
&ACT-DR4   &--   & -- & -- &$-1.18^{+0.40}_{-0.55}$ &--&--\\[2ex]

$\Lambda\text{CDM}+\sum m_{\nu}$ &  Planck&      --   & -- & -- &-- &$<0.107$&--\\
&ACT-DR4   &--   & -- & -- & -- &$<1.47$&--\\[2ex]

$\Lambda\text{CDM}+\alpha_s$ &  Planck&      --   & -- & -- &-- &--&$-0.0055^{+0.0044}_{-0.0067}$\\
&ACT-DR4   &--   & -- & -- &-- &--&$0.060\pm0.028$\\[2ex]

\hline
  
$w\text{CDM}+\Omega_{k}$ &  Planck&        --   & -- & $-0.046^{+0.039}_{-0.012}$ &$-1.30^{+0.94}_{-0.47} $ &--&--\\
& ACT-DR4    &--   & -- & $-0.029^{+0.049}_{-0.009}$ &$-0.84^{+0.73}_{-0.31}$ &--&--\\[2ex]

$\Lambda\text{CDM}+\Omega_{k}+\sum m_{\nu}$ &  Planck&        --   & -- & $-0.077^{+0.041}_{-0.021}$ &-- &$< 0.494 $&--\\
 & ACT-DR4  &--   & -- & $-0.152^{+0.088}_{-0.078}$ &$ -$ &$2.15\pm0.69$&--\\[2ex]

$w\text{CDM}+\Omega_{k}+\sum m_{\nu}$ &   Planck&     --   & -- & $-0.074^{+0.055}_{-0.024}$ &$-1.59^{+0.94}_{-0.75}$ &$0.45^{+0.12}_{-0.37}$&--\\
&ACT-DR4&      --   & -- & $-0.146^{+0.083}_{-0.069}$ &$<-1.30$ &$2.17^{+0.62}_{-0.70}$&--\\[2ex]

$w\text{CDM}+\Omega_{k}+\sum m_{\nu}+\alpha_s$ &  Planck&      --   & -- & $-0.074^{+0.058}_{-0.025}$ &$-1.55^{+1.0}_{-0.75}$ &$0.43^{+0.16}_{-0.37}$&$-0.0005\pm 0.0067 $\\
&ACT-DR4      &--   & -- & $-0.044^{+0.073}_{-0.030}$ &$<-1.06$ &$2.78\pm0.80$&$0.100\pm0.034$\\[2ex]

$w\text{CDM}+\Omega_{k}+\sum m_{\nu}+N_{\text{eff}}$ &  Planck&       --   & $3.04\pm 0.20$ & $-0.074^{+0.056}_{-0.024}$ &$-1.57^{+0.98}_{-0.77}$ &$0.45^{+0.11}_{-0.37}$&--\\
& ACT-DR4    &--   & $2.32\pm 0.52$ & $-0.069^{+0.096}_{-0.041} $ &$-1.6^{+0.8}_{-1.0}$ &$ 1.78\pm 0.66$&--\\[2ex]

$w\text{CDM}+\Omega_{k}+N_{\text{eff}}+\alpha_s$ & Planck&        --   & $2.97\pm 0.24$ & $-0.042^{+0.036}_{-0.012}$ &$-1.30^{+0.89}_{-0.47}$ &--&$-0.0032\pm 0.0081$\\
&ACT-DR4   &--   & $2.8^{+0.7}_{-1.0}$ & $0.014^{+0.046}_{-0.016}$ &$-0.62^{+0.45}_{-0.27}$ &--&$0.085\pm 0.052$\\[2ex]

$w\text{CDM}+\sum m_{\nu}+N_{\text{eff}}+\alpha_s$ & Planck&       --   & $ 2.76\pm 0.22$ & -- &$-1.64^{+0.28}_{-0.40}$ &$ < 0.139 $&$ -0.0098\pm 0.0079$\\
&ACT-DR4     &--   & $3.56\pm0.77$ & -- &$-1.58\pm0.81$ &$2.83\pm0.97$&$0.132^{+0.054}_{-0.034}$\\[2ex]

$w\text{CDM}+\Omega_{k}+\sum m_{\nu}+N_{\text{eff}}+\alpha_s$ & Planck&        --   & $3.03\pm 0.24$ & $ -0.076^{+0.060}_{-0.025}$ &$-1.51^{+0.96}_{-0.72}$ &$< 0.553$&$-0.0004\pm 0.0084$\\
&ACT-DR4   & --   & $3.75\pm 0.80$ & $-0.050^{+0.075}_{-0.034}$ &$ -1.5^{+1.1}_{-0.8}$ &$2.97\pm 0.88$&$0.129^{+0.049}_{-0.035}$\\
\hline
\bottomrule
    \end{tabular}}
    \caption{Constraint at 68\% CL on the extended model parameters for Planck and ACT-DR4.}
\label{tab.Results.Params}
\end{table*}

In~\autoref{tab.Results.SusStat} we summarize the results obtained following the statistical method outlined in the previous section, while in~\autoref{tab.Results.Params} we provide the numerical constraints on the additional parameters included in our MCMC analysis, both for Planck and ACT. 

Notice that within the standard $\Lambda$CDM cosmological model, we recover essentially the same results already discussed in~\cite{Handley:2020hdp}, quantifying the global tension between the two experiments at the level of $2.5\,\sigma$. This should be regarded as the starting point of our investigation where we would like to address the following question: “is there an extension able to accommodate (or even reduce convincingly) this tension?”.

By applying the same methodology to the different cosmological models listed in~\autoref{tab.Results.SusStat}, we observe that all the cases analyzed in this Letter are largely unable to fully solve the global tension between the two datasets. In particular, in some cases, the tension is even increased with respect to the baseline scenario while in most models the disagreement is in fact reduced, but never in a definitive or convincing way, remaining always above 2 standard deviations, see also~\autoref{fig:sigma}. 

The only exception in which the disagreement between ACT-DR4 and Planck is reduced below the threshold of $2\sigma$, is the minimal extension $\Lambda\text{CDM}+N_{\rm eff}$ where the effective number of relativistic degrees of freedom, $N_{\rm eff}$, can vary freely. In this case, our analysis confirms previous results discussed in literature \citep{ACT:2020gnv} about a mild-to-moderate preference of the ACT-DR4 data for smaller amounts of radiation in the early Universe than expected in the Standard Model of particle physics\footnote{In the case of three active massless neutrinos, the Standard Model of particle physics predicts $N_{\rm eff}=3.044$~\citep{Mangano:2005cc,deSalas:2016ztq,Akita:2020szl,Froustey:2020mcq,Bennett:2020zkv,Archidiacono:2011gq} while larger (smaller) values are possible if additional (less) relativistic degrees of freedom are present in the early Universe (see, \textit{e.g.},~\cite{DiValentino:2011sv,DiValentino:2013qma,DiValentino:2015wba,Giare:2020vzo,Giare:2021cqr,DEramo:2022nvb,Baumann:2016wac,Gariazzo:2015rra,Archidiacono:2022ich,An:2022sva} and the references therein).} ($N_{\rm eff}=2.35^{+0.40}_{-0.47}$ at 68\% CL). However it is interesting to point out that this parameter can partially reduce the disagreement between the two experiments at the Gaussian equivalent level of 1.8 standard deviations. 

While such a reduction is clearly not significant enough to claim this model  fits the measurements provided by two most constraining CMB observations better than $\Lambda$CDM, it is worth noting that all the other minimal one-parameter extensions perform worse, with the tension always ranging between $2.9\,\sigma$ (\textit{i.e.}, increasing with respect to the baseline case) and $2.3\,\sigma$ (\textit{i.e.}, only slightly reducing the discrepancy). In particular, allowing a non-standard lensing amplitude $A_{\rm lens}\ne 1$ in the cosmological model, we do not observe great changes in the consistency between the two experiments and estimate a global tension of about $2.6\,\sigma$, close to the $\Lambda$CDM result. On the other hand, considering the possibility of non-flat background geometries or a non-standard Dark Energy equation of state, the tension between Planck and ACT-DR4 slightly reduces to $\sim 2.3\,\sigma$. Conversely, other extensions involving the mass of neutrinos and the running of the spectral index of primordial inflationary perturbations generally increase the inconsistency to about $2.9\,\sigma$. Indeed, as also pointed out by previous studies~\citep{ACT:2020gnv,DiValentino:2021imh,Forconi:2021que}, a comparison of the Planck and ACT-DR4 angular power spectra shows discrepancies at about the upper limit on total neutrino mass and the value of the running of the spectral index, with the ACT-DR4 data preferring larger masses~\citep{ACT:2020gnv,DiValentino:2021imh} and non-vanishing $\alpha_s$~\citep{ACT:2020gnv,Forconi:2021que}, see also~\autoref{tab.Results.Params}. Since these two parameters are only weakly correlated with the other six standard parameters, their tensions are just summed with those of the baseline case, increasing the $\chi^2$ by~\autoref{eq:chi2} and worsening the general agreement between the two experiments.

Along with the minimal extensions discussed so far, we study also many higher-dimensional cosmological models with two or more additional parameters. Here we hazard the hypothesis that increasing the degrees of freedom of the sample can guarantee more freedom in the theoretical model to fit the data, possibly representing a naive way to accommodate the anomalies observed in the CMB angular power spectra, reduce the global disagreement between the experiments, and suggest a new "concordance model". This is why we also analyze very large parameter-spaces with up to 11 free degrees of freedom. However from the results displayed in~\autoref{fig:sigma} and listed in~\autoref{tab.Results.SusStat} it is evident that the disagreement between ACT-DR4 and Planck is not solved in any  such models and we end up in a situation very similar to that described for the minimal extensions with the Gaussian equivalent tension now ranging between $2\,\sigma$ and $3.5\,\sigma$, depending on the specific combination of parameters. In particular, the best "concordance model" we find in this case is a phantom closed scenario with a varying neutrino sector preferring massive neutrinos (see~\autoref{tab.Results.SusStat}), where the global tension is reduced to $2.02\sigma$ (see~\autoref{fig:sigma}).

We therefore conclude that the general disagreement between the Atacama Cosmology Telescope and the Planck satellite is hard to accommodate to below $1\sigma$ by naively extending the cosmological model or by allowing additional parameters to vary. 
The best "concordance model" that our analysis seems to suggest is a minimal 7 parameter $\Lambda$CDM+$N_{\rm eff}$ scenario, where $N_{\rm eff}<3.04$, i.e. the value expected for three active massless neutrinos, implying for example that our current model of e.g. some low-temperature reheating~\citep{DeBernardis:2008zz,deSalas:2015glj}, may be able to lower the global tension below the threshold of $2\sigma$.
Therefore, this "CMB tension" may indicate the standard model of the Universe provides an incorrect or incomplete description of Nature, and our analysis can suggest that a satisfactory solution could require a more radical change in the theory, see for instance~\cite{DiValentino:2020zio, Jedamzik:2020zmd,CANTATA:2021ktz,DiValentino:2021izs,Perivolaropoulos:2021jda,Renzi:2021xii,Schoneberg:2021qvd,Abdalla:2022yfr,DiValentino:2022fjm} and the discussion therein. In addition, it is plausible that significant unaccounted-for systematics in the data are producing biased results in one or both experiments 
and clearly only independent high-precision CMB temperature and polarization measurements such as CMB-S4~\citep{CMB-S4:2016ple,Abazajian:2019eic,CMB-S4:2022ght}, the Simons Observatory~\citep{SimonsObservatory:2018koc,SimonsObservatory:2019qwx}, CLASS~\citep{CLASS}, LiteBIRD \citep{LBIRD}, CORE~\citep{CORE:2016npo,CORE:2017oje}, PICO~\citep{NASAPICO:2019thw} together with forthcoming astrophysical probes and experiments such as Euclid~\citep{EUCLID:2011zbd,EuclidTheoryWorkingGroup:2012gxx}, DESI~\citep{DESI:2013agm,DESI:2016fyo}, The Roman Space Telescope~\citep{Eifler:2020vvg,Eifler:2020hoy} and Rubin Observatory~\citep{Blum:2022dxi} could provide a definitive answer; see also~\citep{CMB-HD:2022bsz,Sehgal:2020yja,Kollmeier:2019ogo,Chluba:2019kpb,Rhodes:2019lur,DiValentino:2020vhf,Chang:2022tzj,Blum:2022dxi} for recent reviews.

\section*{Acknowledgements}

EDV is supported by a Royal Society Dorothy Hodgkin Research Fellowship. WG and AM are supported by "Theoretical Astroparticle Physics" (TAsP), iniziativa specifica INFN. This article is based upon work from COST Action CA21136 Addressing observational tensions in cosmology with systematics and fundamental physics (CosmoVerse) supported by COST (European Cooperation in Science and Technology).
We acknowledge IT Services at The University of Sheffield for the provision of services for High Performance Computing.

\section*{Data Availability}

All the data used are explained in the text and are publicly available.



\bibliographystyle{mnras}
\bibliography{example.bib} 



\bsp	
\label{lastpage}
\end{document}